\documentclass[fleqn,10pt]{wlscirep}
\usepackage[utf8]{inputenc}
\usepackage[T1]{fontenc}

\usepackage[english]{babel}
\usepackage[T1]{fontenc}

\usepackage{ulem}

\usepackage{setspace}

\usepackage{amsmath}
\usepackage{graphicx}
\usepackage[colorinlistoftodos]{todonotes}
\usepackage{hyperref}

\newcommand{\set}[1]{\{#1\}}

\newcommand{\eref}[1]{(\ref{#1})}

\newcommand{\Works}{\Omega}
\newcommand{\swork}{\omega}
\newcommand{\Allworks}{\Works}
\newcommand{\Zeta}{\Works}
\newcommand{\Nu}{\mathrm{N}}

\renewcommand{\P}{\Pi}
\newcommand{\PO}{\P_{\Works}}
\newcommand{\pr}{\pi}
\newcommand{\cw}{\gamma}
\newcommand{\prior}{\alpha_0}
\newcommand{\lo}{\log}
\newcommand{\compl}[1]{\overline{#1}}
\newcommand{\zsize}{m}
\newcommand{\seq}[1]{\{#1\}}
\newcommand{\cnt}{z}

\newcommand{\nov}{\nu}

\newcommand{\infc}{\eta}
\newcommand{\sworkC}{\swork}
\newcommand{\zC}{z_{\compl{\swork}}}

\newcommand{\np}{\nov_P}
\newcommand{\nh}{\nov_H}

\title{Quantifying Novelty and Influence, and the Patterns of Paradigm Shifts}

\author[1]{Doheum Park}
\author[1]{Juhan Nam}
\author[1,2,*]{Juyong Park}
\affil[1]{Graduate School of Culture Technology, Korea Advanced Institute of Science \& Technology, Daejeon, 34141, Republic of Korea}
\affil[2]{Sainsbury Laboratory, University of Cambridge, Cambridge, CB4 1YE, United Kingdom}

\affil[*]{juyongp@kaist.ac.kr}

\begin{abstract}
Recent advances in the quantitative, computational methodology for the modeling and analysis of heterogeneous large-scale data are leading to new opportunities for understanding of human behaviors and faculties, including the manifestation of creativity that drives creative enterprises such as science. While innovation is crucial for novel and influential achievements, quantifying these qualities in creative works remains a challenge.  Here we present an information-theoretic framework for computing the novelty and influence of creative works based on their generation probabilities reflecting the degree of uniqueness of their elements in comparison with other works. Applying the formalism to the data set of a high-quality, large-scale classical piano compositions represented as symbolic progressions of chords--works of significant scientific and intellectual value--spanning several centuries of musical history, we find that the enterprise's developmental history can be characterized as a dynamic process of the emergence of dominant, paradigmatic creative styles that define distinct historical periods. These findings can lead to a deeper understanding of innovation, human creativity, and the advancement of creative enterprises.
\end{abstract}
\begin{document}

\flushbottom
\maketitle
%
%
\thispagestyle{empty}

\section*{Introduction}
Stories of how creative enterprises--science, technology, and art being principal examples--have evolved are often filled with tales of revolutionary, triumphant ``Eurekas'' that usher in a new era: Einstein's theory of relativity, Kekul\'e's determination of the structure of benzene, Tesla's invention of the alternating current (AC) motor, and Brunelleschi's invention of the linear perspective in art are widely-cited examples~\cite{boden2004creative}.  But recent studies have discovered that in reality the evolution of a creative enterprise is driven by innovations--achievements based on new ideas and practice--on many `scales' of significance~\cite{kuhn2012structure,strumsky2015identifying,ackerman1962theory}, rather than only by those that become parts of a legend or folklore.  In order to understand this important phenomenon properly, we must ask why innovations are so valued in human society, and what are the characteristic patterns of their emergence and impact. Recent scientific studies offer some clues. First, studies on human and animal brains have found their innate preference for new stimuli~\cite{bunzeck2006absolute,wittmann2008striatal}.  Second, data-driven studies on citation networks and impacts of scientific papers and patents have shown that novelty is often a key feature in influential scientific knowledge and technological systems~\cite{uzzi2013atypical,kim2016technological}.  This is also true in cultural creations such as music, where continual experimentations of musical elements (e.g. notes, chords, rhythm, etc.) and compositional rules (e.g., modes, scales, forms, tonality, etc.) have long been recognized as instrumental for innovation throughout history~\cite{meyer1957meaning,meyer1989style}. To achieve further progress in answering those questions quantitatively and to fully grasp the role of innovation in the advancement of a creative field, however, there still remains the challenge of how to quantify the novelty of a creative work.  The ability to do so could be very useful in identifying novel works and how a creative form develops dynamically in time. In this work we introduce a foundational information-theoretic framework for computing the novelty of a creative work utilizing its mathematical representation as a set of correlated elements and the past, prior works with which it is compared. We also show that it allows us to define the influence of a creative work on later ones, allowing us to identify the most followed or referenced works that could be understood as having laid the groundwork of the styles found among the works of any given time.

\subsection*{Model and Formalism}
\begin{figure}[ht]
\centering
\includegraphics[width=0.7\linewidth]{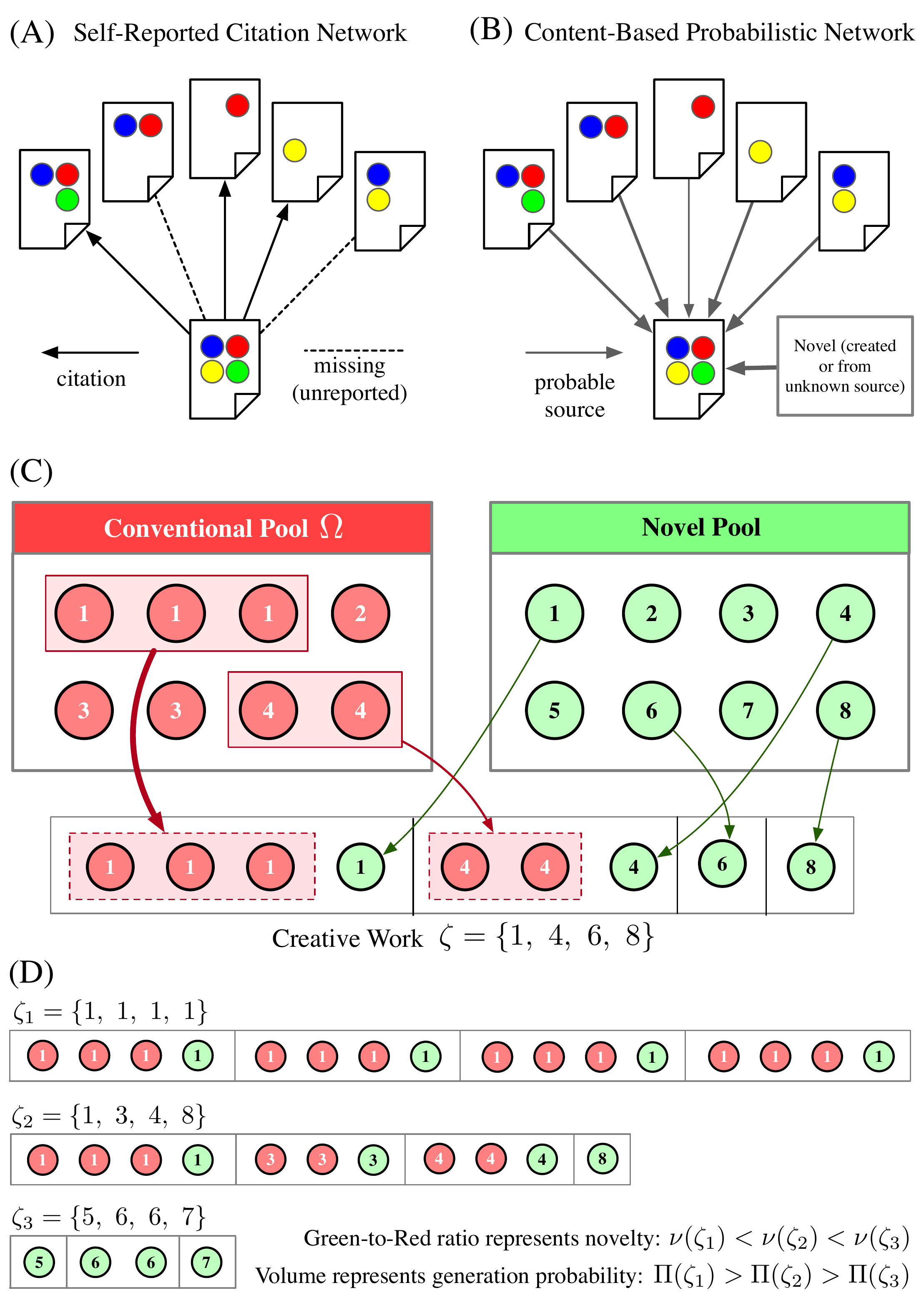}
\caption{Networks of influence and formalism for computing the novelty of a creative work. (A) In a common citation network, a new creation is connected to older ones via self-reported links (solid). Various factors such as human error and bias may cause links to be missing (dotted).  (B) In a content-based probabilistic network, two works that share common elements are viewed as being in a probable influence relationship, therefore no known source is omitted due to error or bias.  Elements that have been truly created or from unknown sources are also accounted for by the so-called Novel Pool, mathematically given as priors.  (C)  Calculating the generation probability of a work $\zeta$ given a set of past works $\Allworks$. The generation process of $\zeta$ is modeled as choosing the elements from the Conventional Pool (CP) that represents `referencing' any of $\Allworks$ or the Novel Pool (NP) that represents `inventing' the element (or unknown sources). CP contains all elements in $\Allworks$ with duplication. Setting up NP to contain a fixed number of each possible element corresponds to the Maximum A Priori estimator, Eq.~\eref{prmap}. The probability of choosing an element is proportional to its count in the combined pool. In $\zeta=\set{1,4,6,8}$, for instance, element $1$ is the most common (four copies total, three from CP and one from NP), whereas $6$ and $8$ are the least so (one from NP only). (D) The total number of copies of a work's elements represents the work's generation probability, whereas the ratio between green (novel) and red (conventional) represents its novelty.  Three examples $\zeta_1$, $\zeta_2$, and $\zeta_3$, all composed of four elements, are compared. The green-to-red ratio represents the novelty $\nu$, while the volume represents the generation probability $\Pi$.}
\label{fig_formalism}
\end{figure}

In order to compute the novelty of a creative work, we first consider the fact that any new creative work--be it a scientific paper, a technological patent, or a musical composition--contains the familiar, `conventional' elements that can be found in known older works, and the unfamiliar, `novel' elements that have not~\cite{uzzi2013atypical,kim2016technological,ackerman1962theory}.  Intuitively then a work that features a larger novel-to-conventional ratio of elements could be considered more novel, and vice versa. How can one tell if an element is conventional or novel? In some form of creative works, notably research publications and patents, the information is explicitly given in the form of a reference or a citation, represented as the solid lines (arrows) in Fig.~\ref{fig_formalism}~(A). The study of citation networks with creative works as nodes and the citations as directed edges with the adjacency matrix
\begin{align}
	\mathbf{A}\equiv\{A_{ij}\}
\end{align}
where $A_{ij}=1$ when paper $i$ is cited by $j$, and $0$ otherwise is a much-studied type in network science~\cite{price1976general,newman2001scientific1,newman2001scientific2,uzzi2013atypical}.  The citation network defined in this manner can be unreliable and incomplete, however, since it relies solely on self-reporting by the creator that is subject to human shortcomings such as faulty memory and bias that could cause missing (unreported) citations, as visualized in Fig.~\ref{fig_formalism}~(A). Also, such citation information, incomplete as it may be, is rarely provided in many other types of written works as well as most cultural works including literature, music, and art. A more reliable method would be then to make a direct comparison between an older work and a new one to identify common elements. If there exists one, it would be an indication that the older work may have been referenced in the creation of the new one. We say only `may have been' because the shared element could have been taken from a different work (known or unknown to the present us), or been `invented' by the creator oblivious of a previous usage. This multiplicity of possible sources suggests a \emph{probabilistic} model of reference that assigns the probabilities that an element in a work has been taken from a known older work, an unknown (lost) older work, or been invented.  This is depicted in Fig.~\ref{fig_formalism}~(B) where the set of known works $\Omega$ is accounted for as a probable source of the generation of $\zeta$, whereas the other possibilities are labeled as the `Novel' source (as they are new to us). We later show that the latter can be implemented mathematically as an uninformed prior.  Fig.~\ref{fig_formalism}~(C) shows in detail how an example work $\zeta=\{1,4,6,8\}$ can be originating from the known works (constituting the `Conventional Pool' of elements) or the novel sources (constituting the `Novel Pool' of elements). For instance, elements common in the conventional pool (such as $1$ and $4$) raise the conventionality (and lower the novelty) of $\zeta$, whereas rarer or nonexistent elements (such as $6$ and $8$) raise its novelty (or lower its conventionality). This is consistent with the intuitive meaning of conventional and novel as being common and familiar (conventional) and rare and unfamiliar (novel). As further examples, in Fig.~\ref{fig_formalism}~(D) we compare three hypothetical works $\zeta_1$, $\zeta_2$, and $\zeta_3$ where the Green-to-Red ratio of elements represents each work's novelty $\nu(\zeta_1)<\nu(\zeta_2)<\nu(\zeta_3)$ which is in the opposite order of the generation probabilities $\Pi(\zeta_1)>\Pi(\zeta_2)>\Pi(\zeta_3)$ represented by the volume under the generation process of Fig.~\ref{fig_formalism}~(B).

Formally, from our model of Fig.~\ref{fig_formalism} we can represent the generation probability $\P_{\Works}(\zeta)$ of a creative work $\zeta=\{e_1,e_2,e_3,\ldots,e_\zsize\}$ as the probability of choosing its elements from the element pools given by
\begin{align}
	\P_{\Zeta}(\zeta)=\prod_{i=1}^\zsize\pr_{\Zeta}(e_i),
\label{pzeta}
\end{align}
where $\pr_{\Works}(e_i)$ is the selection probability of element $e_i$. Since a smaller generation probability $\P$ means that a work is less expected and therefore more novel, the novelty of the work is be a decreasing function of $\P$. We thus define the novelty $\nov(\zeta)$ as the log inverse $\P_{\Zeta}$, normalized by the work $\zeta$'s length $\zsize$ (we take $\log$ to mean $\log_{10}$ in this work):
\begin{align}
	\nov(\zeta) &\equiv \frac{1}{\zsize}\log\frac{1}{\PO(\zeta)} = \frac{1}{\zsize}\sum_{i=1}^{\zsize}\log\frac{1}{\pr_{\Zeta}(e_i)}.
\label{logcon}
\end{align}
This form shows a clear connection to information theory. In information theory, the log of inverse probability of an event is called its information content that measures the unexpectedness (degree of surprise) of an event~\cite{mackay2003information} (measured in bits, had we used log of base 2). Therefore, the novelty is defined to be the average unexpectedness of the elements in the work (the normalization necessary because without it any new work can be made arbitrarily highly novel by lengthening it), in agreement with its intuitive meaning.

Although we have above argued for the appeal of novel works to living beings and their value for progress, it is unlikely that novelty alone is a sign that the work is of any value; if it were, one could simply assemble elements not found in the older works, and claim to have created the most valuable work. In addition to being different from the past, a useful way of gauging a work's value would be to find how much influence a work has had on the posterity, in other words how much a later work has referenced it thereby being directly affected by it. It turns out that we can again use Eq.~\eref{pzeta} to define such \textbf{influence} of an earlier work or a set of earlier works $\swork$ (for instance, the works by a specific creator) on $\zeta$, in other words how strongly $\swork$ has affected the creation of $\zeta$.  Intuitively, we can suspect influence of an earlier work when $\zeta$ shares common elements with it. We say only `suspect' because, as before, the shared elements could have been taken from a different work or invented (i.e. come from the novel pool) by $\zeta$'s creator.  What we can be more certain of, on the other hand, is the \emph{lack} of influence when no element of $\zeta$ is shared with $\swork$, meaning that the elements of $\zeta$ can only be found in  $\overline{\swork}\equiv\Works-\swork$ or the novel pool.  This prompts us to interpret the \emph{difference} between the full generation probability $\PO(\zeta)$ and the reduced one $\P_{\overline{\swork}}(\zeta)$ as $\swork$'s influence on $\zeta$. In other words, influence is the share of the generation probability of $\zeta$ that $\omega$ is accountable for. More specifically, we define $\eta_{\swork}(\zeta)$ as follows, to be consistent with Eq.~\eref{logcon}:
\begin{align}
	\infc_{\swork}(\zeta) &\equiv \frac{1}{\zsize}\log\frac{\P_{\Works}(\zeta)}{\P_{\compl{\swork}}(\zeta)}\ge 0.
\label{influence}
\end{align}
The equality $\infc_{\swork}(\zeta)=0$ (no influence) holds when $\swork$ shares no elements with $\zeta$, i.e. has not contributed at all to its generation.

With these quantitative measures of novelty and influence of creative works, we tackle the following important questions regarding the advancement of creative enterprises: How do the novelty and influence change over time? How do they correlate, i.e. does novelty lead to influence and later recognition? How do these characterize the evolution patterns of a creative field? We illustrate our methodology by analyzing a representative example of a creative enterprise with a long history of high scientific and intellectual value, western classical piano music.

\section*{Novelty and Influence in Classical Piano Music}
We study western classical piano music from the so-called common practice period (circa 1700--1900) chosen for the following advantages: High scientific and cultural significance, widely credited for having produced many fundamental musical styles that are influential today; A rich body of musicological understanding available from traditional research that could be compared with new, alternative approaches such as ours; And the abundance of high-quality data. The availability of large-scale musical databases and advances in scientific, analytical methods continue to enable novel and interesting findings on their properties~\cite{meyer1989style}. Recent examples include researches on the topology and dynamics of the networks of musicians for the discovery of human and stylistic factors in the creation of music~\cite{gleiser2003community,de2004complex,bae2014network,park2014network,park2015topology,bae2016multi} and  stylometric analyses of music that lead to corroborations or fresh challenges to established musicological understanding~\cite{levitin2012musical,serra2012measuring,liu2013statistical,wu2015bach,mauch2015evolution}.

\begin{figure}[ht]
\centering
\includegraphics[width=1.0\linewidth]{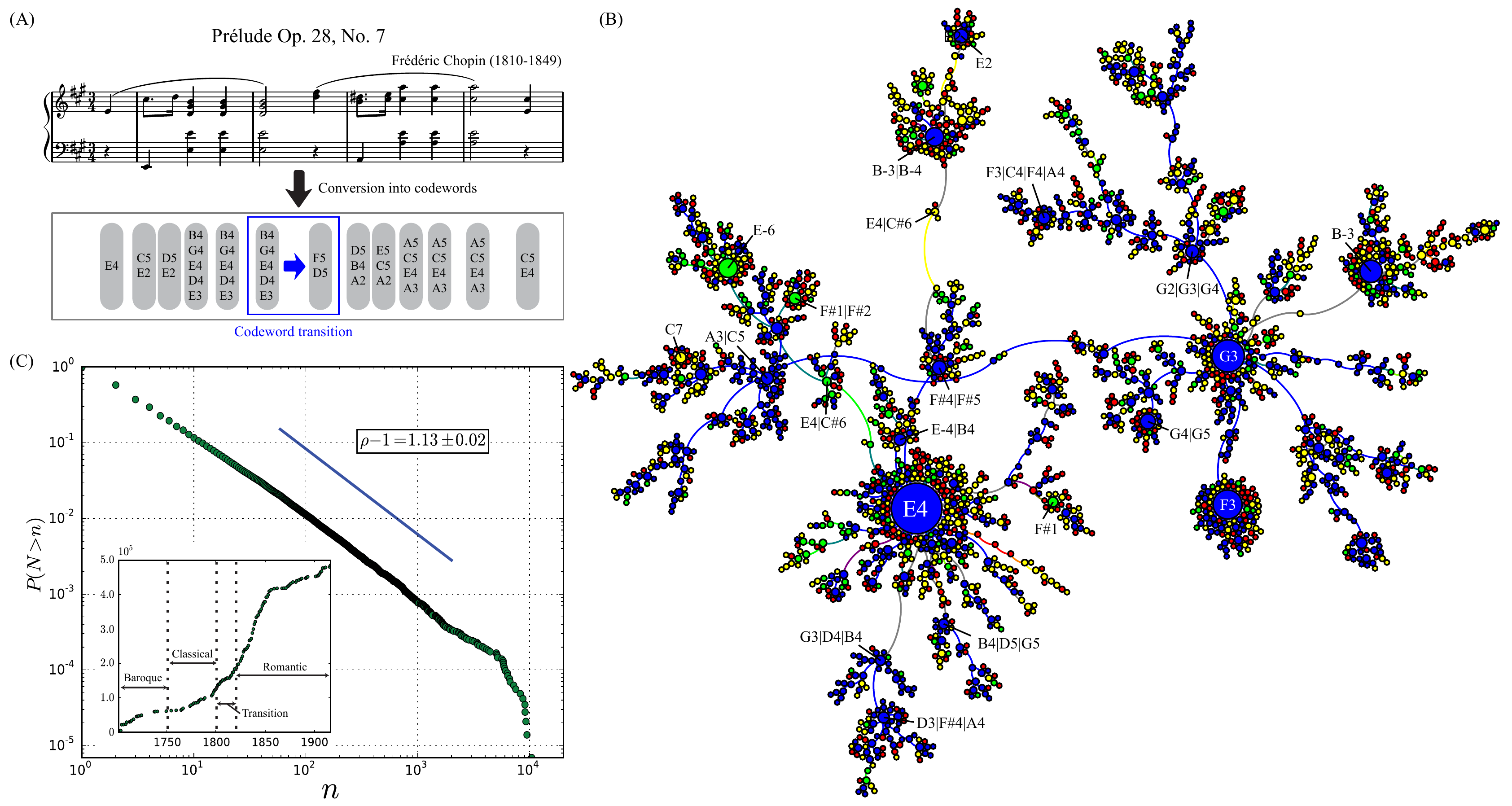}
\caption{(A) A musical score can be converted to a sequence of codewords, simultaneously-played notes, composed of codeword transitions (blue box). (B) The backbone of the network of codeword transitions from our data. Only 2\,267 out of 144\,183 codewords (1.5\%) are shown. The node radius indicates the number of transitions into and out of the corresponding codeword, while the edge width indicates the number of the corresponding transition. The node color indicates the period when the corresponding codeword first appeared (blue-Baroque, green-Classical, yellow-Transition, red-Romantic). (C) The cumulative distribution of the occurrences of the codewords and the cumulative number of unique codeword transitions ever used (inset). The distribution exhibits a highly-skewed, power law-like behavior with power exponent $\rho=2.13 \pm 0.02$ which is found to have been established early in history (Fig S1).}
\label{fig_codeword}
\end{figure}

Using our framework we start by computing the level of novelty in musical compositions and composers, and study how they relate to the known characteristics of music at a given point in history. We then compute their influence on later times and how they can be used to characterize the evolution of compositional styles throughout history. The first step in using the formalism of Eqs.~\eref{logcon}~and~\eref{influence} is representing music as a set of elements, in other words modeling.  Since modeling a system is an abstraction process that necessarily leaves out some real features of the system, it is ideal to retain the most sensible, relevant ones that also suit the modeler's interests.  For instance, for written works such as literature, scientific publications, etc., they could be words or groups of words such as the $n$-grams~\cite{uzzi2013atypical,griffiths2004finding,kim2016technological}, and for paintings they could be colors and shapes~\cite{kao2012computational,kim2014large}.  Here we model a musical composition as a temporally ordered set of simultaneously played nodes or \textbf{codewords}. For the element we take \textbf{codeword transition}, the bigram (2-gram) of codewords. They are shown in Fig.~\ref{fig_codeword}~(A) with the beginning of one of Chopin's Preludes as an example. Otherwise we are leaving out the tempo, rhythm, instrumentation, etc., primarily based on the importance of harmony and melody in the western classical music tradition~\cite{powell2010music} and the fact that for this paper we will be studying one instrument, the piano. Those other elements could be addressed in future research as necessary.  We also note that our definition of a codeword retains all the original information on octaves and the keys in which the works were composed, resulting in a more truthful representation than the one given in Ref.~\cite{serra2012measuring} where only the pitch class was considered (i.e. discarding the octave information; for instance, F4 and F5 were considered both F) and the keys were normalized to C scales. 

Our data set consists of MIDI (Musical Instrument Digital Interface) files collected from Kunst der Fuge (www.kunstderfuge.com) and Classical Piano MIDI (www.piano-midi.de) archives of 900 classical piano works by 19 prominent composers from the common practice period spanning the Baroque (c. 1700--1750), Classical (c. 1750--1820), Classical-To-Romantic Transition (c. 1800--1820), and Romantic (c. 1820--1910) periods, featuring Johann S. Bach and Georg F. Handel of the Baroque era, and Maurice Ravel of the late Romantic era.  The composers and their works are in the supplementary material (SI Dataset 2). The MIDI files were converted into musicXML format via MuseScore2 software and chordified using Music21, a python library toolkit for computer-aided musicology~\cite{cuthbert2010music21}. The \texttt{chordify} method in Music21 converts a multiple-part complex musical score into a series of simultaneous notes as visualized in Fig.~\ref{fig_codeword}~(A). Since each codeword transition is a directed dyad, they can be collectively visualized as a network whose backbone is shown in Fig.~\ref{fig_codeword}~(B). The cumulative distribution of the number of occurrences of the codewords is shown in Fig.~\ref{fig_codeword}~(C), and approximates a power law with exponent $\rho=2.13 \pm 0.02$, indicating significant disparities in popularity between codeword transitions. Although such a pattern is established early in history (Fig. S1), the number of unique codeword transitions ever used also constantly increases in time (inset of Fig.~\ref{fig_codeword}~C), with the highest rate of increase observed during the Romantic period.

We now compute the novelty and influence of musical compositions. Writing a composition $\zeta$ as a sequence of codewords $\zeta=\seq{\cw_1,\cw_2,\ldots,\cw_\zsize}$ the generation probability of $\zeta$ is given by the first-order Markov chain
\begin{align}
	\P_{\Zeta}(\zeta) &= \pr_\Zeta(\cw_1)\pr_{\Zeta}(\cw_1\to\cw_2)\cdots\pr_{\Zeta}(\cw_{\zsize-1}\to\cw_{\zsize}),
\label{pZ}
\end{align}
For $\pr_\Zeta$ we employ the Maximum A Priori (MAP) estimator~\cite{murphy2002learning} commonly used in Markov chains, given as
\begin{align}
	\pr_\Zeta(\cw_i\to\cw_j) = \frac{\cnt(\cw_i\to\cw_j)+\prior(\cw_i\to\cw_j)}{\sum_{\cw\in\Gamma}\bigl(\cnt(\cw_i\to\cw)+\prior(\cw_i\to\cw)\bigr)},
	\label{prmap}
\end{align}
where $\cnt(\cw_i\to\cw_j)$ is the number of the $\cw_i\to\cw_j$ transition in the conventional pool $\Zeta$ and $\prior(\cw_i\to\cw_j)$ is the prior representing the novel pool in our scheme.  When it is constant it is called the uninformed prior, and we set $\alpha_0\equiv1$ so that our novel pool contains exactly one copy of each possible transition. $\Gamma$ is the codeword space.  The probability of the first codeword $\pr_\Zeta(\cw_1)$ is similarly $\pr_\Zeta(\cw_1) = (\cnt(\cw_1)+1)/(\sum(\cnt(\cw)+1))$, where $\cnt(\cw_1)$ is the number of occurrences of $\cw_1$ as the first codeword in $\Zeta$. Plugging this into Eq.~\ref{logcon}, we obtain the novelty
\begin{align}
	\nu(\zeta)	&\equiv \frac{1}{\zsize}\log\frac{1}{\P_\Zeta(\zeta)} =\frac{1}{\zsize}\biggl[\lo\frac{1}{\pr_\Zeta(\cw_1)}+\sum_{k=1}^{\zsize-1}\lo\frac{1}{\pr_\Zeta(\cw_k\to\cw_{k+1})}\biggr].
	\label{def:novelty}
\end{align}

\subsection*{Historical and Psychological Novelty}
When computing the novelty of Eq.~\eref{def:novelty}, we are free to choose $\Zeta$, the reference set of previous works that determine the conventional pool.  A straightforward choice of $\Zeta$ would be all known works that preceded $\zeta$ in history. This was aptly given the name \textbf{historical novelty}  (H-novelty) by Artificial Intelligence (AI) research circles~\cite{boden2004creative}, and represents a given work's novelty within the entire history of the field up to its creation.  Another interesting choice of $\Zeta$ contains all the previous works by the very creator of $\zeta$.  The resulting novelty is named \textbf{psychological novelty} (P-novelty)~\cite{boden2004creative} that represents, for instance, the degree of improvement in a new version of an algorithm or a machine over its previous versions.  Applied to our data it would show how a composer evolves in compositional style against his own past works~\cite{margulis2014repeat}.

\begin{figure}[ht]
\centering
\includegraphics[width=0.6\linewidth]{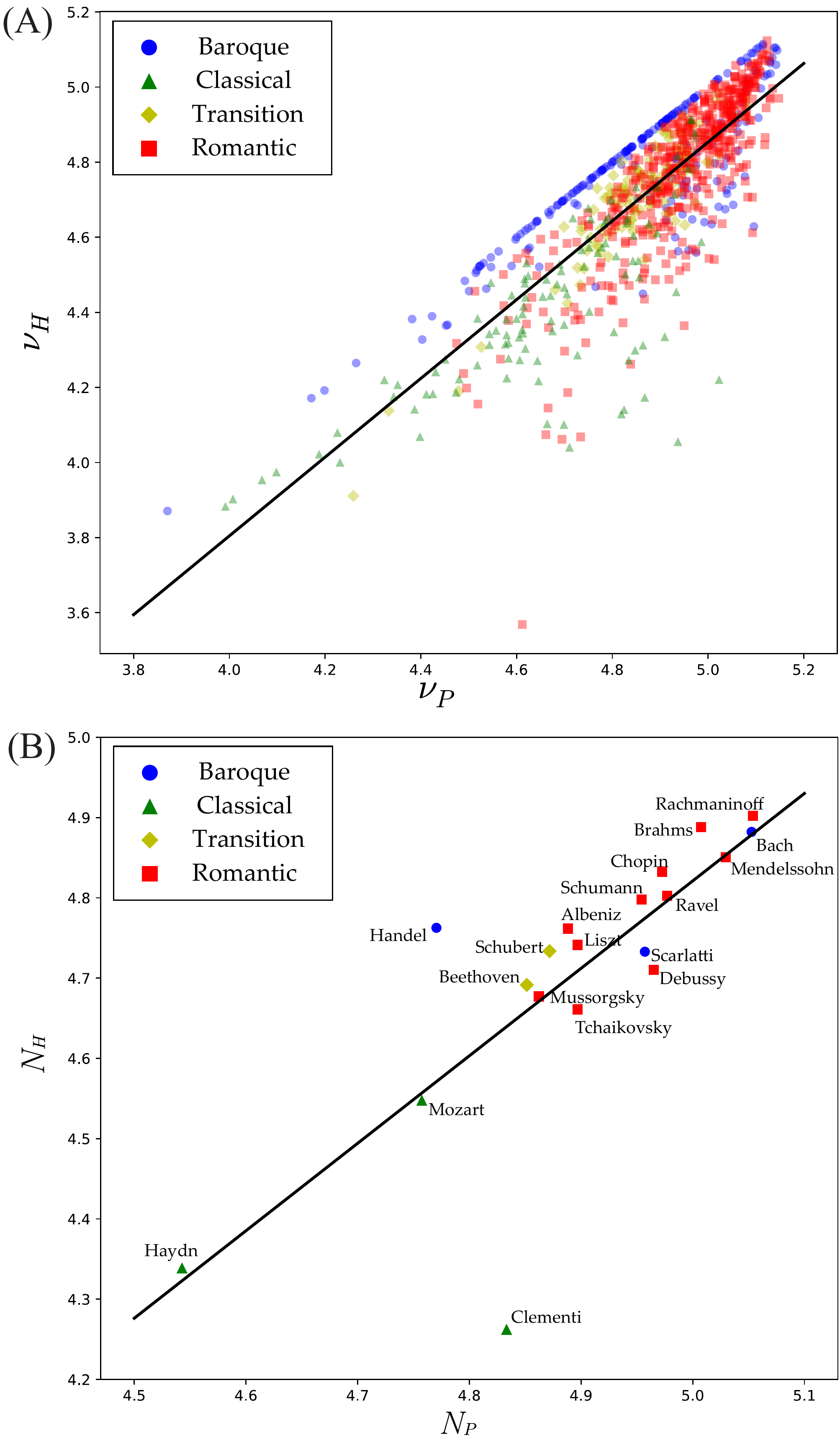}
\caption{(A) Scatterplot of the H- and P-novelty scores of the piano works, with a high level of correlation (Spearman correlation $0.82 \pm 0.01$). (B) The H- and P-novelty scores of composers (Spearman correlation $0.83 \pm 0.11$). A notable outlier is Clementi who shows a lower H-novelty than his P-novelty.}
\label{fig_p_and_h_novelty_correlation}
\end{figure}

We show in Figs.~\ref{fig_p_and_h_novelty}~(A)~and~(B) the cumulative distributions of the H- and P-novelties of the piano works in our data for each period. Of the four, the Classical compositions tend to score low in both novelties, showing that many past conventions were reused both historically and psychologically (see Fig. S2 for the H- and P-novelty scores of the pieces over time).  The novelties of the composers (given by the average of their works') noted $\Nu_H$ and $\Nu_P$ are shown in Figs.~\ref{fig_p_and_h_novelty}~(C)~and~(D).  We note that our confidence in the high H-novelty of the Baroque composers is low due to the much smaller conventional pool than other periods.  For the same reason, however, the raised H-novelty for the Romantic composers should be considered more impressive since it is achieved against the largest conventional pool.  The high level of P-novelty shows Romantic composers having also actively introduced diverse and new codeword transitions throughout their careers.  This is in clear agreement with the widely-accepted thesis that credits Romantic composers with having broken many accepted musical conventions and having diligently conducted personal experimentation of new combinations of pitches~\cite{kravitt1992romanticism}.  The H- and P-novelties are generally positively correlated throughout, with the Spearman correlations equal to $0.820 \pm 0.013$ for the compositions and $0.827 \pm 0.113$ for the composers, respectively, meaning that pursuing novelty involved deviating from both the others and oneself (Fig.~\ref{fig_p_and_h_novelty_correlation}).  The most notable outlier from this trend is Muzio Clementi whose H-novelty is significantly lower than his P-novelty would suggest, as shown in Fig.~\ref{fig_p_and_h_novelty_correlation}~(B). This means that while he produced works distinct from his earlier works (even more so than Handel, Mozart, and Haydn, and on par with Beethoven), they as a whole would sound conventional when compared with others'. This may form a quantitative corroboration for the the common assessment of Clementi that in his time his reputation rivaled Haydn's among his contemporaries, but languished for much of the 19th century and beyond~\cite{youngren1996finished}: The diversity of codeword transitions that he employed in his works (reflected in the high P-novelty) could have been the source of high reputation among his contemporaries, then as time passed his works failing to distinguish themselves from other traditional works (reflected in the low H-novelty) could have caused his loss in stature.

\subsection*{Influence and Shifts in Dominant Styles}
While novel achievements are indispensable for the progress and growth of a creative enterprise, our results above suggest that novelty alone would not cause one to be considered `the greatest'; Beethoven, for instance, stand among the lower half in computed novelty.  This is in line with many recent research findings that a creative work's impact on its posterity does not depend solely on the degree of its novelty, and how it builds on tradition is also important~\cite{uzzi2013atypical,kim2016technological,ackerman1962theory}.  Musical composition would be no exception: Past works exert influence on the future by serving not only as training material for new composers, but also by inspiring new works or themselves being tweaked and transformed into new original works~\cite{meyer1989style,boden2004creative}.  Even mimicry or imitation, normally associated with subpar works lacking in originality and artistic value, can sometimes occur in renowned masters' works and gain recognition: Franz Liszt, a leading Romantic-era composer, admired Beethoven so much that in a famous act of homage he transcribed Beethoven's complete symphony cycle into the piano~\cite{searle1980liszt} that is now considered a significant and influential achievement in its own right. These observations tell us that the definition of `influence' of a work as the degree to which it has been referenced by later works as in Eq.~\eref{influence} is a sensible one.

To compute $\infc_{\swork}(\zeta)$ of Eq.~\eref{influence}, the influence of composer $\swork$ on $\zeta$, we start by rewriting $z(\cw_i\to\cw_j)$, the number of $\cw_i\to\cw_j$ transitions in $\Zeta$, in Eq.~\eref{prmap} as
\begin{align}
	z(\cw_i\to\cw_j) = z_{\swork}(\cw_i\to\cw_j)+\zC(\cw_i\to\cw_j),
\end{align}
where $z_\sworkC$ is the number of instances of the transition used by $\swork$, and $z_{\compl{\swork}}$ is that by all the other composers before $\zeta$. Then $\P_\Zeta(\zeta)$ becomes
\begin{align}
	\P_\Zeta(\zeta) &= \frac{\bigl(z_\swork(\cw_1)+\zC(\cw_1)+1\bigr)}{\sum_{\cw\in\Gamma}\bigl(z(\cw)+1\bigr)} \times  \frac{\bigl(z_\swork(\cw_1\to\cw_2)+\zC(\cw_1\to\cw_2)+1\bigr)}{\sum_{\cw\in\Gamma}\bigl(z(\cw_1\to\cw)+1\bigr)}\times\cdots.
\label{infgen1}
\end{align}

Eliminating all $z_\swork$s in the numerator, we obtain
\begin{align}
\Pi_{\compl{\swork}}(\zeta) &= \frac{\bigl(\zC(\cw_1)+1\bigr)}{\sum_{\cw\in\Gamma}\bigl(z(\cw)+1\bigr)} \times \frac{\bigl(\zC(\cw_1\to\cw_2)+1\bigr)}{\sum_{\cw\in\Gamma}\bigl(z(\cw_1\to\cw)+1\bigr)}\times\cdots
\label{infgen2}.
\end{align}

\begin{figure}[ht]
\centering
\includegraphics[width=1.0\linewidth]{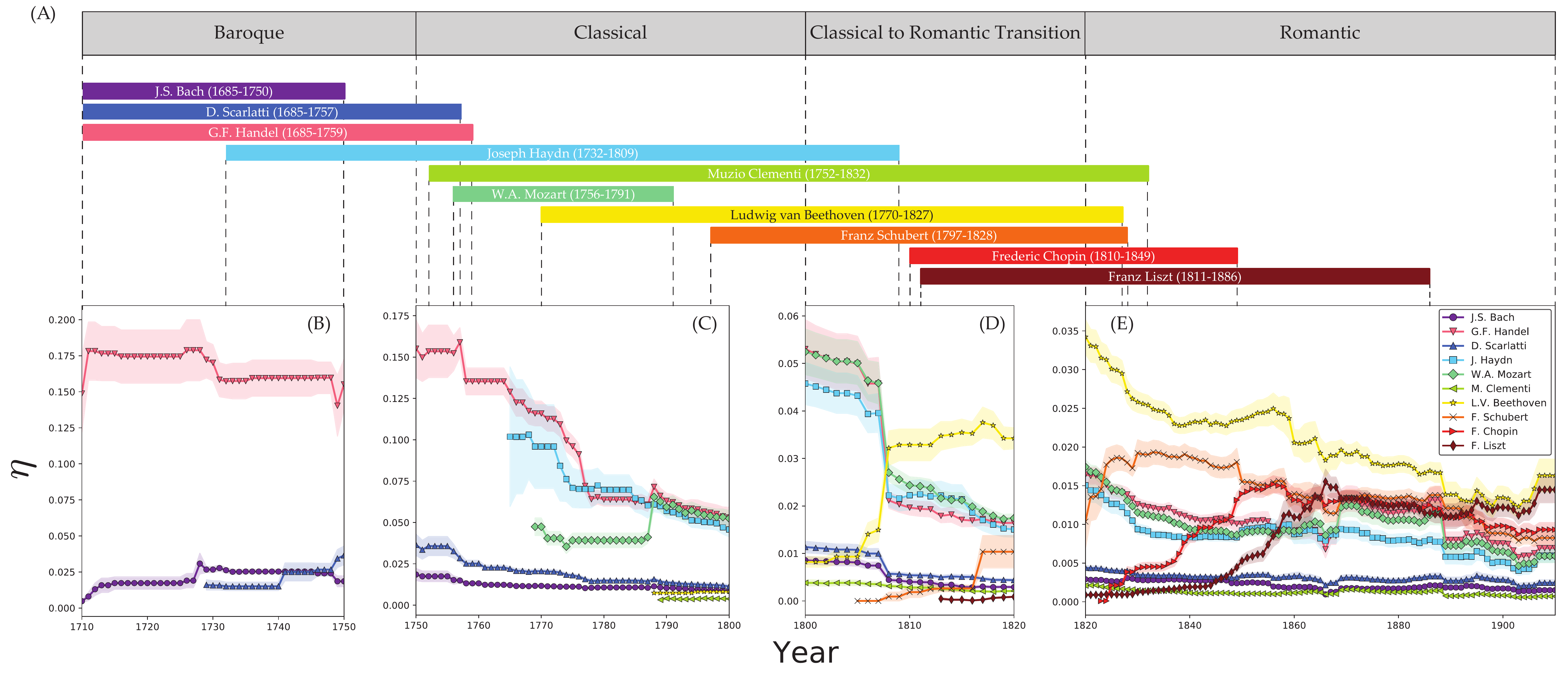}
\caption{The influence of composers. (A) The common period designations and the living years of ten major composers in our data. (B)--(E) The mean influence $\eta$ of the composers on the works composed at a given time $t~(\pm 10~\textrm{years})$. Each period is distinguished by the emergence of newly dominant composers that indicate paradigmatic shifts in composition styles, and provide quantitative support for period designations. (B) During the Baroque period Handel exerts a dominant influence on other composers. (C) In the Classical period initially Scatlatti's influence increases, while Handel's influence begins to wane. Then Haydn and Mozart's influence rival Handel's. (D) The Classical-to-Romantic Transition period is characterized by Beethoven who overtakes the most influential ones from the previous period (Handel, Haydn, and Mozart). (E) The Romantic period also witnesses the emergence of newly highly influential composers such as Schubert, Chopin and Liszt.}
\label{fig_influences}
\end{figure}

After computing the influences $\set{\eta}$ between all $7\,298$ eligible composer--composition pairs (self-influences were excluded) we plot each composer's mean influence on the works created at any given time $t$ ($\pm10$ years for smoother curves), shown in Fig.~\ref{fig_influences}.  During the Baroque period (B) Handel is the most influential, indicating that his codeword transitions were often also used at a later time by his contemporaries Bach and Scarlatti, whereas the opposite did not occur as frequently.  More interesting patterns can be found when we observe the rise and fall of the composers' influences over time. Since a high influence means that later works share common elements, we can interpret such rise and fall of composers' influence as indicating the shifts in compositional style, and providing a quantitative justification for the distinct period labels. Let us examine, as a start, the Baroque and the Classical periods in Figs.~\ref{fig_influences}~(B)~and~(C).  While Handel maintains his dominant influence until around the mid-Classical period, we identify two notable patterns: First, Scarlatti overtakes Bach in influence shortly before the Classical period, in agreement with the well-acknowledged significance of Scarlatti on the Classical period~\cite{taruskin2009music}; Second, Haydn and Mozart emerge during the Classical period with a high influence, soon rivaling Handel's. Similar dynamics--the rise or emergence of a new leading influential figure--are observed in subsequent periods. The Classical-to-Romantic transitional period (Fig.~\ref{fig_influences}~D) is characterized by the emergence of Beethoven whose historical significance~\cite{burkholder2006history} is clearly shown. Beethoven's high influence in this period shows his younger contemporaries adopting his codewords more willingly than any other predecessor's (Figs.~S4~C~and~D) that continues well into the Romantic period. Then during the Romantic period new composers such as Schubert, Chopin, and Liszt rise in influence to rival or overtake Mozart and Beethoven (Fig.~\ref{fig_influences}~E), befitting their reputation as of finally eclipsing those ``classical sounds'' and establishing many essential repertoire now permanently associated with the piano~\cite{burkholder2006history}.

\section*{Discussion}
\begin{figure}[ht]
\centering
\includegraphics[width=1.0\linewidth]{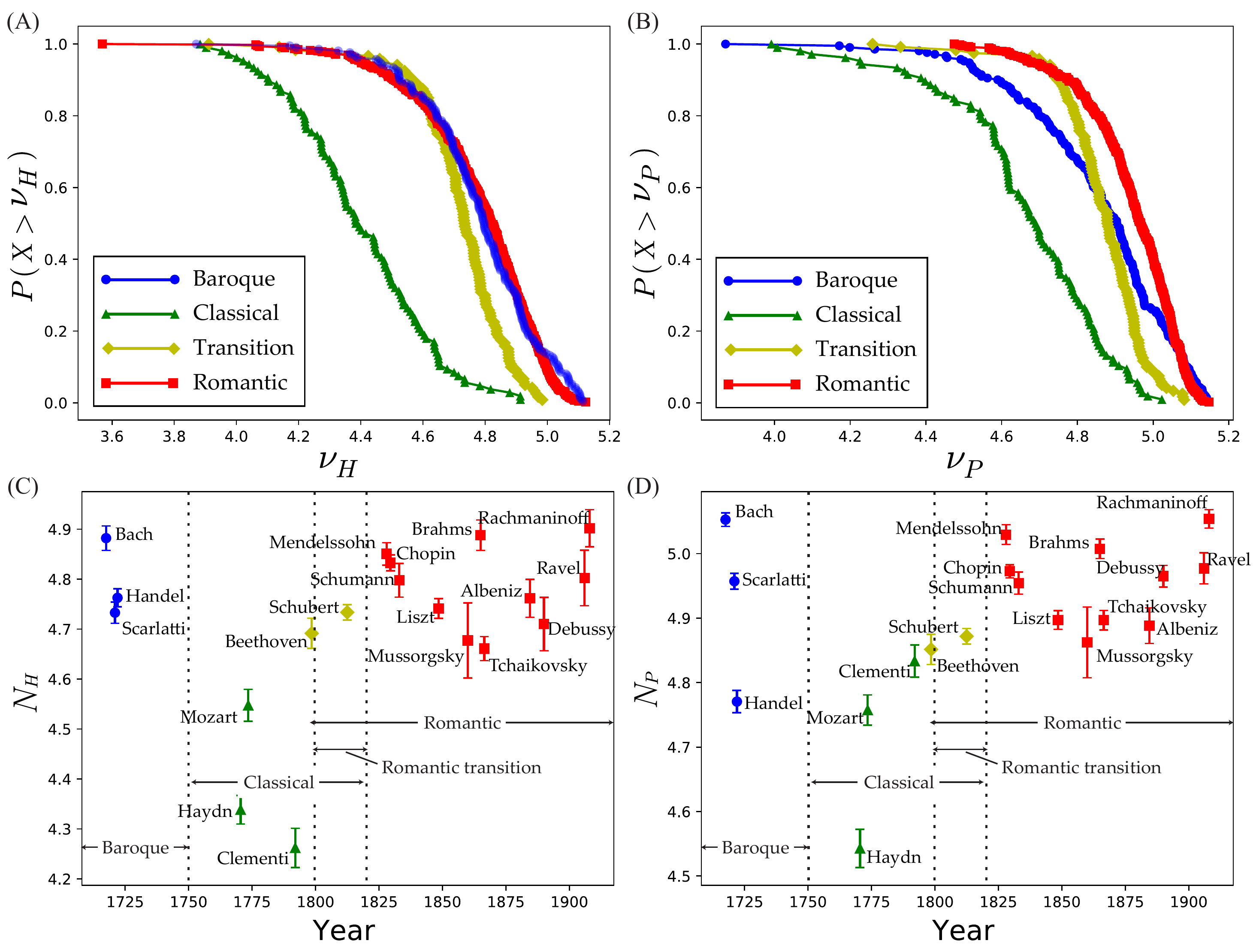}
\caption{The H-novelty (left) and P-novelty (right) scores of the piano works (top) and the composers (bottom). (A) The cumulative distribution of the H-novelty scores $\nh$ of the works. The median and mean values are $(4.80, 4.78)$ for the Baroque, $(4.38, 4.40)$ for the Classical, $(4.73, 4.729)$ for the Transition, and $(4.82, 4.78)$ for the Romantic periods. (B) The cumulative distribution of the P-novelty scores $\np$ of the works. The median and the means are $(4.90, 4.86)$ for the Baroque, $(4.69, 4.66)$ for the Classical, $(4.88, 4.87)$ for the Transition, and $(4.97, 4.94)$ for the Romantic periods. (C) \& (D) The novelty $\Nu_H$ and $\Nu_P$ of the composers (defined as the mean of $\nh$ and $\np$ of their works). A composer's position on the $x$-axis (year) is the midpoint between his birth and death years.}
\label{fig_p_and_h_novelty}
\end{figure}

This work presents a general mathematical framework for computing the novelty and influence of creative works based on the degree of shared elements between past and future works. Novelty measures how different a work is from the past, representing originality and unpredictability of generation. Influence measures how much a work has been referenced in the future, representing its success and impact as an inspiration for future creations. While originality and success are both important characteristics of meaningful creative works, they do not correlate perfectly. Handel was less novel than Bach and many others but had more influence on Classical and Romantic composers (Figs.~\ref{fig_p_and_h_novelty}~and~\ref{fig_influences}) is a good example.  Similarly, Beethoven, Schubert, and Liszt were less novel than Mendelssohn and Schumann (Fig.~\ref{fig_p_and_h_novelty}), but eventually came to exert more influence and inspire more piano music to follow (Fig. S4).  The separation between novelty and influence is particularly auspicious in the case of music from the Classical period (especially Mozart): Mozart is shown to have used fewer novel codewords per se and opted to use the conventions from the Baroque period, but his works nevertheless had enough high artistic value that he gained much influence in the future.  This is another example of our analysis agreeing with traditional musicology that identifies the Classical period as ``valuing of shared conventions, rational restraint and the playful exploitation of established constraints''~\cite{meyer1989style}. This contrasts with the composers of the later Romantic period who introduced new elements in a faster pace (Fig.~\ref{fig_p_and_h_novelty}) but again agrees with the traditional musicological assessment of their efforts in ``pursuing the value of being individual, peculiar and original.''~\cite{meyer1989style,taruskin2010music,burkholder2006history}

We note that, while we employed the simplest first-order Markov model of codeword transitions to model music, the framework is general enough for a higher-order Markov model or related techniques such as Hidden Markov Model (HMM) and neural networks that have been previously applied for analyzing text and music~\cite{sutskever2011generating,graves2013generating,kim2016character,murphy2002learning,rabiner1986introduction,gers1999learning,mikolov2010recurrent}. An application of higher-order Markov models shows a broad agreement with our main findings using first-order Markov. One notable extra finding made possible by using higher-order Markov involves Debussy whose greatest innovation  is believed to have been in the use of non-traditional scales. (Figs. S5, 6, 7, 8, 9, 10 and 11)

We finally discuss the potential issues of using curated data such as ours and how they are addressed in our framework. One can justifiably point out that throughout the common practice era considered in our analysis there existed many active composers not included in our data set but who nevertheless likely referenced and influenced one another. Note that this situation is not unique to our data, but is becoming increasingly common in an era when interesting data are collected from many open real-world systems where one cannot easily expect them to be as complete or comprehensive as those from designed experiments conducted in highly-controlled laboratory environments. It is more important and practical, then, to deal with the situation by incorporating appropriate mechanisms in the methodology and understanding the nature of the data. First, in our formalism, incomplete data results in underrepresented or missing elements in the Conventional Pool. But our formalism addresses this issue via the Novel Pool that gives some weight to unknown or as yet unknowable cases via the uninformed prior, a well-established, unbiased method employed in statistics in the absence of usable information. Additionally, they can be updated whenever new information becomes available in a straightforward manner.  Second, our data comprises works that are at the time of this study the most highly regarded, and most often studied and performed, implying that of all imaginable data sets of similar size it would be among the most commensurate with the meaning of novelty and influence: Since they are by definition based on what is available for comparison, our data would be closer to a modern listener's true experience than others comprising obscure or less popular works would be, rendering it both desirable and useful given the conditions. 

\section*{Conclusion}
The availability of a quantitative computational methodology confers the ability to confirm or challenge existing knowledge and understanding about a system in a statistically robust manner, and to find more detailed and advanced answers to long-standing or new questions.  In this paper we proposed a framework for quantifying the novelty of a creative work and the influence between those produced at different times based on the intrinsic compositions of the works.  As an example, we applied it to the development of classical music by using 900 classical piano compositions that cover the common practice era in the western musical tradition. As the intrinsic element of music we focused on the codeword transitions to measure the novelty and influence of the compositions and composers. From the use of codeword transitions over time, we found that commonly designated ``periods'' corresponded well to the emergence of newly influential composers indicating notable shifts in styles. In addition to a broad agreement with conventional understanding of the characteristic of periods and composers, an interesting finding was that being more novel, i.e. more willing to break from convention, did not necessarily translate to being influential on the posterity.  This means while novelty is still necessary in a creative endeavor--high-novelty composers in our data set are undoubtedly universally recognized masters of the form themselves--it cannot account for all the creative, artistic qualities that facilitated those codeword transitions that were more widely transmitted to later generations.

This suggests a future research direction in which a more elaborate modeling of codeword transitions and other elements of music are considered. Possibilities in the former category include the change of the number of notes~\cite{meyer1989style}, the tonality~\cite{cuddy1995expectancies,temperley2007music}, melody~\cite{narmour1991top,schellenberg1996expectancy,zivic2013perceptual} and the chord progression~\cite{loui2007harmonic}, to name a few.  Possibilities in the latter category include the rhythmic structure of music that is recently gaining increased attention~\cite{levitin2012musical,margulis2014repeat} and the global structure of a composition, given the common assertion that the most significant innovation in the piano music during the Classical period was the establishment of the sonata form~\cite{burkholder2006history} which may have little to do with the codeword transitions. Extension beyond the piano is also an obvious possibility, as many composers we considered were prolific in other forms including Haydn who is also very well known for his chamber music and symphonies~\cite{webster2003new}. 
  
Given the generality of our methods, we envision our framework proposed here being useful in addressing many questions pertinent to the development of various cultural, creative fields and genres other than music as we have presented here. We believe such scientific approach to the subject will permit a new level of understanding of human creativity and eventually shed more light on the dynamics of the progress of intellectual or cultural products. 

\bibliography{ref}

\section*{Acknowledgements}
The authors thank Kyungmyeon Lee for helpful discussions. This work was supported by the National Research Foundation of Korea (NRF-20100004910, NRF-2013S1A3A2055285 and NRF-2016S1A2A2911945) and the BK21 Plus Postgraduate Organization for Content Science.

\section*{Author contributions statement}
DP, JP conceived and designed the experiments. DP, JP performed the experiments. DP, JN, JP analyzed the data. DP, JP contributed reagents/materials/analysis tools. DP, JN, JP wrote and reviewed the manuscript. DP crawled data.

\section*{Additional information}
\textbf{Competing interests} The author(s) declare no competing interests.

\section*{Data Availability}
All data generated or analysed during this study are included in this published article and its Supplementary Information files.

\end{document}